\documentstyle[12pt,damtp]{damtpprint}
\pubdate{September 1991}
\pubnumber{DAMTP-91/31}
\title{Singular Vectors of the Virasoro Algebra}
\author{Adrian Kent\damtp}
\abstract{
We give expressions for the singular vectors in the highest
weight representations of the Virasoro algebra.
We verify that the expressions --- which take the form of a product
of operators applied to the highest weight vector --- do indeed define
singular vectors.
These results explain the patterns of embeddings amongst Virasoro algebra
highest weight representations.
}

\begin{document}
\maketitle

\input mssymb
\newenvironment{proof}{{\bf Proof}}{}
\newtheorem{lemma}{Lemma}
\newtheorem{theorem}[lemma]{Theorem}
\newtheorem{corollary}[lemma]{Corollary}
\def\floor#1{\lfloor#1\rfloor}
\def\blank#1{}
\def\swap#1#2{#1}
\def \ul {\underline}
\def \eps {\epsilon}
\def\ut#1{\hbox{\boldmath #1}}
\def\cO{{\cal O}}
\def\tr{{\rm tr}}
\def\romif{{\rm~if~}}
\def\sgn{{\rm sgn}}
\def\sig{{\rm sig}}
\def\diag{{\rm diag}}
\def\for{{\rm ~for~}}
\def\and{{\rm ~and~}}
\def\romor{{\rm ~or~ }}

Conformal field theory relies on a description of the Virasoro algebra's
highest weight representations, and in particular on the classifications
of the levels at which representations have singular vectors\cite{kac,ff1}
and the embedding relations amongst these vectors.\cite{ff2,ff3}
These embedding relations are encoded in the
irreducible Virasoro characters,\cite{rocha} from which the partition
functions of conformal field theories are built.\cite{bpz,cardy,ciz}

While this information is enough for most conformal field theoretic
purposes, there are several applications\cite{mattis,langlands,ak1} in which
explicit expressions for the singular vectors are needed.
The first relevant work was the beautiful paper of Malikov, Feigin and
Fuchs,\cite{mff} which gives expressions for the general
singular vectors in finite dimensional Lie algebra Verma modules.
Feigin and Fuchs\cite{ff3} have also presented some partial results describing
projections, and asymptotic properties, of the Virasoro algebra singular
vectors.
Benoit and Saint-Aubin\cite{bsa} (BSA) found remarkable explicit expressions
for the
sub-class of the singular vectors $v_{p,q}$ in which either $p$ or $q$ is $1$.
Recently, Bauer et al.\cite{bdfiz1,bdfiz2} have rewritten the BSA
expressions in a
compact form in which their singularity is manifest, and in which very
interesting connections to integrable systems and to W-algebra theory
appear.  Bauer et al. have also given a new algorithm by which any vector
$v_{p,q}$ can in principle be calculated: we shall discuss this later.
In this letter, we give expressions for all the singular vectors
$v_{p,q}$, show how these expressions explain the
embeddings of the Virasoro algebra's highest weight representations, and
sketch proofs of these results.

First let us recall some basic facts and describe the results of Benoit and
Saint-Aubin.
The Virasoro algebra has commutation relations
\eqn{vcrs}
\begin{array}{rcl}
{}~[L_m , L_n ] \, \, & = & (m - n ) L_{m+n} + \frac{m^3 - m}{12}
\delta_{m,-n} C \, , \\ ~[L_m , C ] \, \, & = & 0 \, .
\end{array}
\en
The Verma module $V(h,c)$ is the representation which
contains a vector $\ket{h}$ such that
\eqn{hwt}
\begin{array}{rcll}
L_m \ket{ h } &=& 0 & \romif m>0 \, , \\ L_0 \ket{ h } &=& h
\ket{h} \, , & \\ C \ket{h} &=& c \ket{h} \, , &
\end{array}
\en
and which has a basis comprising the states
$L_{-i_1} \ldots L_{-i_r} \ket{h}$ with
\mbox{$i_1 \geq \ldots \geq i_r > 0$.}
We have the decomposition
\eq
V(h,c) = \bigoplus_{n = 0,1,2 \ldots}  V_{n} (h,c)  \, ,
\en
where the level $n$ space
$V_n (h,c ) $ is the eigenspace of $L_0$ with eigenvalue $(h+n)$.
Define a singular vector in $V(h,c)$ to be a vector $v$, lying in some
$V_n (h,c)$ for $n>0$, with the property that
\eqn{sing}
L_m v = 0  \romif m>0 \, .
\en
It is known\cite{kac,ff1,ff2,ff3} that there is a
singular vector at level $N$ in
$V(h,c)$ if and only if, for some positive integers $p$ and $q$ and complex
number $t$, we have $N = pq$, and
\eqn{hc}
\begin{array}{rclcl}
c &=& c(t) &=& 13 - 6 t - 6 t^{-1} \, , \\
h &=& h_{p,q}(t)   &=& {\displaystyle
  \frac{p^2 -1}{4} t - \frac{pq -1}{2} + \frac{q^2 -1}{4} t^{-1}} \, .
\end{array}
\en
It is also known that the singular vector at level $N$, when it exists, is
unique up to scalar multiplication.
Thus, given $p$ and $q$, the singular vector $v_{p,q}$ is a
function of $t$.  In fact, one can
write $v_{p,q}(t) = O_{p,q} (t) \ket{h_{p,q}(t)}$,
where
\eq
O_{p,q} (t) = \sum_{|I| = pq} a^{p,q}_I (t) L_{-I}
\en
and the $a^{p,q}_I (t)$ depend polynomially on $t$ and $t^{-1}$.
Here the sum is over sequences
$I = \{ i_1 , \ldots , i_n \}$ of positive integers ordered so that
$i_1 \geq \ldots \geq i_n $, we write $L_{-I} = L_{-i_1} \ldots L_{-i_n}$
and $|I| = i_1 + \ldots + i_n$, and
take the coefficient of $(L_{-1})^{pq}$ to be $1$.

However, this turns out not to be the most convenient form in which to
describe the operators $O_{p,q}(t)$.
In the cases when $p=1$ or $q=1$,
BSA obtained remarkably simple expressions for the operators:
\eqn{bsaops}
\begin{array}{rcl}
O_{p,1} (t) &=&
{\displaystyle \sum_{{I = \{ i_1 , \ldots , i_n \}}  \atop{ |I| = p}}}
c_p ( i_1 , \ldots , i_n ) (-t)^{p-n} L_{-I} \, , \\
O_{1,q} (t) &=&
{\displaystyle \sum_{{I = \{ i_1 , \ldots , i_n \}}  \atop{ |I| = q}}}
c_q ( i_1 , \ldots , i_n ) (-t)^{-q+n} L_{-I} \, .
\end{array}
\en
These sums are over all sequences of positive integers summing to $p$ or
$q$, without any ordering restriction.
The coefficients are defined by
\eq
c_r ( i_1 , \ldots , i_n ) =
{\displaystyle \prod_{{1 \leq k < r} \atop{k \neq \sum_{j=1}^s i_j
{\rm~for~any~}s}}} k (r - k ) \, .
\en

Our first step in generalising these results follows the
ideas of Malikov-Feigin-Fuchs\cite{mff} by
extending the enveloping algebra of the Virasoro algebra to
include operators of the form $(L_{-1})^{a}$ for arbitrary
complex values of $a$.  Thus, as well as the relations (\ref{vcrs}),
we have
\eqn{vtcrs}
\begin{array}{rcll}
{}~[L_m , ( L_{-1} )^a ] &=& {\displaystyle \sum_{n=1}^{m+1}}
({\displaystyle\prod_{r=1}^n}
\frac{(m+2-r)(a+1-r)}{r}) ( L_{-1} )^{a-n} L_{m-n}
& \romif m \geq 0 \, , \\
{}~[ (L_{-1} )^a , L_m ] &=& {\displaystyle \sum_{n=1}^{\infty}}
({\displaystyle\prod_{r=1}^n }
\frac{-(m+2-r)(a+1-r)}{r})  L_{m-n} (L_{-1})^{a-n}
                                         & \romif m < 0 \, ,
\end{array}
\en
and
\eq
(L_{-1})^a L_{-1} = L_{-1} (L_{-1})^a = (L_{-1})^{a+1} \, , ~~~
(L_{-1})^a (L_{-1})^b = (L_{-1})^{a+b}.
\en
(That is, we are considering the central extension of
the algebra generated by differential operators
$z^n d$ and generalised pseudodifferential operators $d^a$.)
Denote the algebra generated by $L_m , C$ and the $(L_{-1})^a$
by $\tilde{V}$.  Define
the $\tilde{V}$ representation $\tilde{V}(h,c)$ to be a generalised
Verma module with a
vacuum vector $\ket{h}$ and on which
$C$ acts as the scalar $c$.
That is, equations (\ref{hwt}) hold
and the vectors $L_{-n_1} \ldots L_{-n_r} ( L_{-1} )^a \ket{h}$, with
$n_1 \geq \ldots n_r \geq 2$ and $a$ unrestricted,
form a basis for $\tilde{V}(h,c)$.
Note that $(L_{-1})^a \ket{h}$ is not zero, even when $a$ is negative, so that
$\ket{h}$ is neither highest nor lowest weight in $\tilde{V}(h,c)$.
We shall be interested in the Virasoro singular vectors in $\tilde{V}(h,c)$ ---
that is, those
vectors $v$ lying in some $L_0$ eigenspace and obeying equation (\ref{sing}).

We say an infinite sum in $\tilde{V}$ is a well-defined operator if it has the
form
\eqn{wd}
a_0 (L_{-1})^a + \sum_I a_I L_{-I} (L_{-1})^{a - |I|} \, ,
\en
where the coefficients $a_0$ and $a_I$ are finite and the
sum is over sets $I = \{ i_1 , \ldots , i_n \}$ with $n$ finite
and the integers $i_j \geq 2$;
we say that the operator is of level $a$ and that a
term $a_I L_{-I} (L_{-1})^{a - |I|}$ is of order $|I|$.
We also say an operator is well-defined if it can be reduced to the form
(\ref{wd}) by commuting all $(L_{-1})^a$ terms (of any power $a$)
through to the right and if moreover, for any $n$, the operator can be
reduced to the form
\eq
a_0 (L_{-1})^a + \sum_{|I| \leq n} a_I L_{-I} (L_{-1})^{a - |I|} +
O( (L_{-1})^{a - n -1} )
\en
by a finite number of these reordering operations.
Here $O( (L_{-1})^{m} )$ means a sum of
monomials of the form
\eq
A M_1 ( L_{-1} )^{a_1} M_2 ( L_{-1} )^{a_2} \ldots M_r ( L_{-1} )^{a_r} M_{r+1}
\en
where $A$ is a scalar, the $M_i$ are either $1$ or products of $L_{-i}$ with
$i \geq 2$, and $a_1 + \ldots + a_r \leq m$.
This means in particular that if two operators are well-defined then
their product is also well-defined.

The naive generalisation of expressions (\ref{bsaops}) to $\tilde{V}$
is not well-defined.
However, a well-defined generalisation can be obtained in the following way.
First, rewrite the expressions (\ref{bsaops}) by commuting all $L_{-1}$
operators to the right, with no other reordering.  This gives
\eq
\begin{array}{rcl}
O_{p,1} (t) &=& {\displaystyle
\sum_{{k_1 \ldots k_r} \atop{ k_i \geq 2}}}
P_{k_1, \ldots , k_r} (p,t) L_{- k_1} \ldots L_{- k_r}
(L_{-1})^{p - \sum k_i } \, , \\
O_{1,q} (t) &=& {\displaystyle
\sum_{{k_1 \ldots k_r} \atop {k_i \geq 2} }}
P_{ k_1, \ldots , k_r} (q, t^{-1}) L_{- k_1} \ldots L_{- k_r}
(L_{-1})^{q - \sum k_i }
\, ,
\end{array}
\en
where $P_{k_1, \ldots , k_r} (p , t)$ is defined for $p \geq \sum_i k_i$,
in which range it is a polynomial function in $p$ and $t$.
We analytically extend $P_{k_1, \ldots , k_r} (p , t)$ to arbitrary $p$ and
define operators in $\tilde{V}$ by
\eqn{anal}
\begin{array}{rcl}
\cO_{a,1} (t) &=& (L_{-1})^a + {\displaystyle \sum_{n=2}^{\infty}} \,
 {\displaystyle \sum_{{k_1 + \ldots + k_r = n} \atop {k_i \geq 2}}}
P_{ k_1, \ldots , k_r } (a, t) L_{- k_1} \ldots
L_{- k_r} (L_{-1})^{a - n } \, , \\
\cO_{1,b} (t) &=& (L_{-1})^b + {\displaystyle \sum_{n=2}^{\infty}} \,
 {\displaystyle \sum_{{k_1 + \ldots + k_r = n} \atop {k_i \geq 2}}}
P_{k_1, \ldots , k_r } (b , t^{-1}) L_{- k_1} \ldots
L_{- k_r} (L_{-1})^{b - n } \, ,
\end{array}
\en
for any complex numbers $a$ and $b$.
Now, when applied to the vacua, these operators
create Virasoro singular vectors
in the modules $\tilde{V}(h_{a,1}(t), c(t))$ and
$\tilde{V}(h_{1,b}(t), c(t))$.
To see this, consider the vector
\eq
L_1 \cO_{a,1}(t) \ket{ h_{a,1}(t) }
\en
expressed as a sum of the form
\eq
\sum_{k_1 \geq \ldots \geq k_r \geq 2}
Q_{k_1 , \ldots , k_r}(a, t, t^{-1} ) L_{-k_1} \ldots L_{-k_r}
(L_{-1})^{a - k_1 - \ldots - k_r -1} \ket{h_{a,1}(t)} \, .
\en
Each $Q_{k_1 , \ldots , k_r} (a , t, t^{-1} )$ is a polynomial,
obtained as a sum of multiples of polynomials
$P_{ k'_1, \ldots , k'_s } (a, t)$ with
$(\sum_{i=1}^s k'_i ) \leq (\sum_{i=1}^r k_i ) + 2$.
Thus if $N$ is an integer larger than $(\sum_{i=1}^r k_i ) + 1$,
we have that $Q_{k_1 , \ldots , k_r } (N , t, t^{-1} )$ is the coefficient
of
\eq
L_{-k_1} \ldots L_{-k_r}
(L_{-1})^{N - k_1 - \ldots - k_r -1} \ket{h_{N,1}(t)}
\en
in the
expression $L_1 O_{N,1} \ket{h_{N,1}(t)}$.
That is, $Q_{k_1 , \ldots , k_r} (N , t, t^{-1} ) = 0$ for all integers
$N \geq (\sum_{i=1}^r k_i ) + 2$, and so the polynomial $Q$ must be
identically zero.

Similarly, $L_2 \cO_{a,1}(t) \ket{ h_{a,1}(t)}$,
$L_1 \cO_{1,b}(t) \ket{ h_{1,b}(t)}$, and
$L_2 \cO_{1,b}(t) \ket{ h_{1,b}(t)}$ are all zero;
hence $\cO_{a,1}(t) \ket{ h_{a,1}(t)}$ and
$\cO_{1,b}(t) \ket{ h_{1,b}(t)}$ are Virasoro singular vectors.

It is now easy to see that a vector
of the form
\eq
X_{a_n} \ldots X_{a_1} \ket{h} \, ,
\en
where each $X_a = \cO_{a,1} (t)$ or $\cO_{1,a} (t)$,
will be singular provided that, for each $r$ from $1$ to $n$,
if $X_{a_r} = \cO_{a_r ,1} (t)$ then $h + \sum_{i=1}^{r-1} a_i = h_{a_r , 1}$
and if $X_{a_r} = \cO_{1, a_r } (t)$ then
$h + \sum_{i=1}^{r-1} a_i = h_{1, a_r }$.

To make use of these expressions, we first need to show that if
$X$ is a well-defined operator of level $0$, and if
$X \ket{h}$ is a Virasoro singular vector in $\tilde{V}(h,c)$, then $X$ is
a scalar multiple of the identity.  Suppose this is not so,
and expand $X$ as a sum over canonically ordered partitions:
\eq
X = a_0 + {\displaystyle
\sum_{{I  = \{ i_r, .... , i_1 \}}\atop {i_r \geq \ldots \geq i_1 \geq 2}}}
a_I L_{-I} (L_{-1})^{|I|} \, .
\en
If $I = \{ i_r, .... , i_2 , i_1 \}$ and $J = \{ j_s , .... , j_2 , j_1 \}$
are two
canonically ordered partitions, let $I > J$ if $|I| > |J|$, and
if $|I| = |J|$ let $I>J$ if for some $k$ we have $i_k > j_k$ and
$i_l = j_l$ for all $l$ with $l<k$.
Then let $I' = \{ i'_r , \ldots , i'_1 \}$ be the lowest partition with
non-zero coefficient in $X$
(that is, $a_{I'} \neq 0$ and if $I < I'$ then $a_I = 0$).
Then, letting $I'' = \{ i'_r , \ldots ,  i'_2 \}$, we have that the
coefficient of $L_{-I''} (L_{-1})^{|I''| - i'_1 +1} \ket{h}$ in
$L_{i_1 -1} X \ket{h}$ is
non-zero, contrary to our original assumption.

Now it follows from equation (\ref{hc}) that
\eq
\begin{array}{rclcl}
h_{a,1} + a &=& h_{a',1}  & \Rightarrow &
       a' = -a \romor a' = a + 2 t^{-1}   \, , \\
h_{1,b} + b &=& h_{1,b'}  & \Rightarrow &
       b' = -b \romor b' = b + 2 t        \, , \\
h_{a,1} &=& h_{1,b}   & \Rightarrow &  t( a \pm 1) = (1 \pm b)    \, .
\end{array}
\en
Hence $\cO_{-a,1} (t)\cO_{a,1} (t) \ket{h}$ is a singular vector at level $0$.
Since
\eq
\cO_{-a,1} (t)\cO_{a,1} (t) = (1 + O( (L_{-1})^{-2})) \, ,
\en
and since
similar
observations apply to $\cO_{1,-b} (t)\cO_{1,b} (t)$, we have that
\eqn{ama}
\begin{array}{rcl}
\cO_{-a,1} (t)\cO_{a,1} (t) &=& 1 \, , \\
\cO_{1,-b} (t)\cO_{1,b} (t) &=& 1 \, .
\end{array}
\en
A similar argument shows that
$\cO_{-p,1} (t)O_{p,1} (t) = 1 = \cO_{1,-q} (t)O_{1,q} (t) $ for
positive integers $p$ and $q$, and so
\eq
\begin{array}{rcl}
\cO_{p,1} (t) &=& O_{p,1} (t)\, , \\
\cO_{1,q} (t) &=& O_{1,q} (t)\, .
\end{array}
\en
That is, the operators (\ref{anal}) are analytic extensions of the BSA
operators (\ref{bsaops}).  (This is not obvious from their
definition.)  It also follows that if \mbox{$h = h_{a,1} (t) = h_{1,b}
(t)$}, with \mbox{$t(1+a) = (1+b)$}, then
\eq
\cO_{-a,1} (t) \cO_{1,-b-2} (t) \cO_{a+2,1} (t) \cO_{1,b} (t) \ket{h}
\en
is a singular vector at level $0$, and so we have the identity
\eqn{comm}
\cO_{1,t(1+a) +1 } (t) \cO_{a,1} (t) = \cO_{a+2,1} (t) \cO_{1,t(1+a) -1} (t)
\, .
\en
Thus, for generic $(h,c)$, the vector $\ket{h}$ in $\tilde{V}(h,c)$
lies at one vertex of a commutative diagram which takes the form of an
infinite rectangular lattice whose points correspond to Virasoro
singular vectors and whose edges correspond to operators of the form
$\cO_{a,1}$ and $\cO_{1,b}$.  (See Figure 1.)

Next we consider $\tilde{V}(h,c)$ when $c= c(t)$ and $h = h_{p,q} (t)$
for some
positive integers $p$ and $q$.  We have
\eq
\begin{array}{rcl}
a = p - (q-1)t^{-1} & \Rightarrow & h_{p,q}(t) = h_{a,1}(t)  \, , \\
b = q - (p-1)t & \Rightarrow & h_{p,q}(t) = h_{1,b}(t)  \, .
\end{array}
\en
Hence the vectors
\eqn{sv1}
\cO_{p+(q-1)t^{-1},1} (t) \cO_{p+(q-3)t^{-1},1} (t) \ldots
\cO_{p-(q-1)t^{-1},1} (t) \ket{h_{p,q}(t)}
\en
and
\eqn{sv2}
\cO_{1,q+(p-1)t} (t) \cO_{1,q+(p-3)t} (t) \ldots
\cO_{1,q-(p-1)t} (t) \ket{h_{p,q}(t)}
\en
are Virasoro singular vectors at level $pq$ in $\tilde{V}(h_{p,q}(t),c(t))$.
But, since
\eq  \cO_{-p+(q-1)t^{-1},1} (t)  \ldots
\cO_{-p-(q-1)t^{-1},1} (t) O_{p,q}(t)
\ket{h_{p,q}(t)}
\en
and
\eq
\cO_{1, -q + (p-1)t} (t) \ldots
\cO_{1, -q - (p-1)t} (t) O_{p,q}(t) \ket{h_{p,q}(t)}
\en
are singular vectors at level $0$, we have that
\eqn{soln}
\begin{array}{rcl}
O_{p,q}(t) &=&
\cO_{p+(q-1)t^{-1},1} (t) \cO_{p+(q-3)t^{-1},1} (t) \ldots
\cO_{p-(q-1)t^{-1},1} (t) \\
&=& \cO_{1,q+(p-1)t} (t) \cO_{1,q+(p-3)t} (t) \ldots
\cO_{1,q-(p-1)t} (t) \, .
\end{array}
\en
Note that, although the individual operators on the
right hand side of equations (\ref{soln})
do not generally belong to the Virasoro
enveloping algebra, the two products do: the equations describe
operator identities, not merely relations that hold to
$O((L_{-1})^{-1})$.
This means that (\ref{sv1}) and (\ref{sv2}) are in fact two equivalent
expressions
for the singular vector $v_{p,q} (t)$ in the Virasoro algebra
Verma module $V(h_{p,q}(t),c(t))$.  These general
formulae for the singular vectors of the Virasoro algebra are our main
results.

An immediate application of these results is an explanation of the
regularity of the Virasoro algebra's Verma module embeddings, and
hence of its character formulae.  For example, take a discrete
series representation $V(h_{p,q}(t), c(t))$ with $t= (m+1)/m$, for some
integers $m,p,q$ with $m \geq 2$ and $1 \leq q \leq p \leq (m-1)$.
Its singular vectors fall into
an embedding pattern of type $III_{-}$ in Feigin-Fuchs'
classification.\cite{ff2,ff3} The regularity of this pattern derives from
two types of identities, of which the simplest examples are
\eq
O_{m+p, m+1+q} (t) = O_{2m+p,q} (t) O_{m+p, m+1-q} (t) O_{p,q} (t)
\en
and
\eq
O_{m+p,m+1-q} (t) O_{p,q} (t) = O_{p,2(m+1)-q} (t) O_{m-p,m+1-q} (t)
\, .
\en
(See Figure 2.)  The first of these follows directly from equation
(\ref{soln}); both sides are equal to
\eq
\prod_{r=0}^{m+q} \cO_{(m+p)+(m+q -2r)t^{-1},1} (t) \, ,
\en
where for non-commuting operators we set $\prod_{i=1}^r A_i = A_1 \ldots A_r$.
The second identity follows from equations (\ref{soln}) and (\ref{comm}).  We
have
\eqn{prod1}
O_{m+p, m+1-q} (t) O_{p,q} (t) =
{\displaystyle \prod_{r=0}^{m-q}}
\cO_{(m+p) + (m-q-2r)t^{-1},1} (t)
{\displaystyle \prod_{r=0}^{p-1}}
\cO_{1,q + (p-1-2r)t} (t)
\en
and
\eqn{prod2}
\begin{array}{l}
O_{p,2(m+1)-q} (t) O_{m-p,m+1-q} (t)  = \\
{}~~~~~~~~~~~~ {\displaystyle \prod_{r=0}^{p-1}}
\cO_{1,(2(m+1) -q) + (p-1-2r)t} (t)
{\displaystyle \prod_{r=0}^{m-q}}
\cO_{(m-p) + (m-q -2r) t^{-1},1} (t) \, .
\end{array}
\en
Then equation (\ref{comm}) shows that the relevant quadrilateral
in figure 2 can be refined to an $\cO$ operator
commutative diagram in the form of a
$p \times (m+1-q)$ rectangular lattice, with
the right hand sides of (\ref{prod1}) and (\ref{prod2}) each
represented by two of the outer edges.

The literature now contains quite a few results about Virasoro singular
vectors, and so we conclude with a brief comparative discussion.
Perhaps it is worth remarking that the Kac determinant formula\cite{kac,ff1}
already provides not only an existence theorem, but also an algorithm for
calculating each singular vector $v_{p,q}(t)$: the singular vector is
the zero eigenvector of the inner product matrix $M_{pq} (h_{p,q}(t),c(t))$
at generic $t$, and the inner product matrix can be obtained from the
Virasoro algebra's commutation relations.
Bauer et al.\cite{bdfiz2} gave recursion relations which define a
chain of vectors of levels $0,1,2, \ldots, pq$ in the Verma module
$V (h_{p,q}(t),c(t))$; the last of these vectors is the singular
vector $v_{p,q}(t)$.
While these results haven't yet led to explicit expressions for the
singular vectors, they might well eventually do so; they might also ---
as Bauer et al. suggest --- lead to a new proof of the Kac determinant
formula, and perhaps also of the Feigin-Fuchs classification.
This would certainly be
valuable: an intrinsic method of proving determinant and character formulae
may well be necessary in understanding the representation theory of
some of the Virasoro algebra's extensions.
In any case, Bauer et al. have described an interesting
algebraic structure within the Verma module, which needs to be better
understood; the connections with W-algebra theory and
integrable models are particularly fascinating.

Our expressions for the Virasoro singular vectors in the modules
$\tilde{V}(h_{p,q}(t),c(t))$ are simple and completely explicit.
One can easily calculate the projection of one of these vectors to the Verma
module $V (h_{p,q}(t),c(t))$.
However, we still do not have general explicit expressions ---
that is, expressions given
solely in terms of elementary functions and enveloping algebra
elements --- for the singular vectors in the Verma modules.
Such expressions would certainly be of some intellectual
interest.  On the other hand, we expect that the product formulae
(\ref{soln}), together with the relations (\ref{ama}) and (\ref{comm}), will
actually prove much more useful in any theoretical applications.
For example, it seems
unlikely that the Verma module embeddings can be understood so readily
without use of the product formulae (\ref{soln}).
(Compare Malikov-Feigin-Fuchs' formulae\cite{mff} for the
$sl(n)$ singular vectors: the Verma module embeddings can easily be
read off from the complex exponent formula (19), but are thoroughly
obscured in equation (20).)

To summarise: the Virasoro singular vectors
derive from a simple algebraic structure within the modules $\tilde{V}(h,c)$,
in which the analytically continued BSA operators are the analogues of
the powers of simple roots used by Malikov-Feigin-Fuchs in the Kac-Moody
case.\cite{mff}
We suspect that similar structures underlie the highest weight
representation theory of the Virasoro algebra's extensions.

\acknowledgement

I am very grateful to P. Goddard,
H. Kausch, G. Watts and J.-B. Zuber for helpful
discussions.  This work was supported by an SERC Advanced
Fellowship and by the Knox-Shaw Research Fellowship at Sidney Sussex
College, Cambridge.

\newpage
\begin{picture}(400,500)(-200,-400)
\put (0,0){\circle{3}}
\put (-15,30){$\ket{h_{a,1}(t)}$}
\put (0,0){\vector(-1,-1){100}}
\put (-75,-50){$\cO_{a,1}$}
\put (55,-50){$\cO_{1,t(1+a)-1}$}
\put (-105,-150){$\cO_{1,t(1+a)+1}$}
\put (60,-150){$\cO_{a+2,1}$}
\put (65,50){$\cO_{-a+2t^{-1},1}$}
\put (-115,50){$\cO_{1,-t(a-1)+1}$}
\put (0,0){\vector(1,-1){100}}
\put (0,0){\vector(-1,1){100}}
\put (0,0){\vector(1,1){100}}
\put (100,-100){\circle{3}}
\put (-100,-100){\circle{3}}
\put (0,-200){\circle{3}}
\put (100,-100){\vector(1,1){100}}
\put (100,-100){\vector(1,-1){100}}
\put (100,-100){\vector(-1,-1){100}}
\put (-100,-100){\vector(-1,1){100}}
\put (-100,-100){\vector(1,-1){100}}
\put (-100,-100){\vector(-1,-1){100}}
\put (0,-200){\vector(-1,-1){100}}
\put (0,-200){\vector(1,-1){100}}
\put (-35,-400){Figure 1}
\end{picture}
\newpage
\begin{picture}(200,400)(-100,-400)
\put (0,0){\circle{3}}
\put (-20,15){$\ket{h_{p,q}(m)}$}
\put (-75,-50){$O_{p,q}$}
\put (55,-50){$O_{m-p,m+1-q}$}
\put (-165,-150){$O_{m+p, m+1-q}$}
\put (-145,-250){$O_{2m+p,q}$}
\put (110,-150){$O_{2m-p,q}$}
\put (110,-250){$O_{3m-p,m+1-q}$}
\put (15,-110){$O_{p,2(m+1)-q}$}
\put (-75,-110){$O_{m-p, m+1+q}$}
\put (0,0){\vector(-1,-1){100}}
\put (0,0){\vector(1,-1){100}}
\put (100,-100){\circle{3}}
\put (-100,-100){\circle{3}}
\put (100,-100){\vector(-2,-1){200}}
\put (-100,-100){\vector(2,-1){200}}
\put (100,-100){\vector(0,-1){100}}
\put (-100,-100){\vector(0,-1){100}}
\put (100,-200){\circle{3}}
\put (-100,-200){\circle{3}}
\put (100,-300){\circle{3}}
\put (-100,-300){\circle{3}}
\put (100,-200){\vector(0,-1){100}}
\put (-100,-200){\vector(0,-1){100}}
\put (100,-200){\vector(-2,-1){200}}
\put (-100,-200){\vector(2,-1){200}}
\put (-101,-305){.}
\put (-101,-310){.}
\put (-101,-315){.}
\put (-101,-320){.}
\put (99,-305){.}
\put (99,-310){.}
\put (99,-315){.}
\put (99,-320){.}
\put (-35,-400){Figure 2}
\end{picture}

\end{document}